\documentclass[twocolumn,twoside,slac_two]{revtex4}
\usepackage{graphicx}
\usepackage{fancyhdr}
\begin{document}

\title{VERITAS Observations of Very High Energy Blazars and Potential for Cosmological Insight}


\author{A. Furniss for the VERITAS Collaboration}
\affiliation{Stanford University, Stanford, CA, 94305 USA}

\begin{abstract}
Gamma-ray blazars are among the most extreme astrophysical sources, harboring phenomena far more energetic than those attainable by terrestrial accelerators. These galaxies are understood to be active galactic nuclei that are powered by accretion onto supermassive black holes and have relativistic jets pointed along the Earth line of sight. The emission displayed is variable at all wavelengths and timescales probed thus far, necessitating contemporaneous broadband observations to disentangle the details of the emission processes within the relativistic jets. The very high energy (VHE; $E\ge$100 GeV) photons emitted by these sources are detectable with ground-based imaging atmospheric Cherenkov telescopes such as VERITAS. As these photons propagate extragalactic distances, the interaction with the diffuse starlight that pervades the entire Universe results in a distance and energy dependent gamma-ray opacity, offering a unique method for probing photon densities on cosmological scales. These galaxies have also been postulated to be potential sources of ultra-high-energy cosmic rays, a theory which can be examined through deep gamma-ray observations of sources which probe moderate gamma-ray opacities. Within this work, I will highlight ongoing research regarding the broadband emission from VERITAS-observed VHE blazars, as well as the potential to use them for cosmological insight.

\end{abstract}

\maketitle

\thispagestyle{fancy}


\section{VERITAS}
VERITAS is an array of four imaging atmospheric Cherenkov telescopes in southern Arizona, each with a 3.5$^{\circ}$ field of view.  The array is sensitive to very-high-energy (VHE) gamma-ray photons with energies from $\sim$70 GeV to more than 10 TeV and can detect a 1\% Crab-Nebula-flux source at 5 standard deviations ($\sigma$) in less than 28 hours.  The telescope array uses 12-meter reflectors to focus dim, blue/UV Cherenkov light from gamma-ray and cosmic-ray interactions in the atmosphere onto cameras composed of 499 photomultiplier tubes (PMTs).  More details on the VERITAS instrument can be found in \cite{holder} and \cite{weekes}.

\section{Studying Cosmology with VHE Blazars}
Blazars are a type of active galaxy with a relativistic jet pointed toward the observer.  These sources are perplexing objects which contain some of the most energetic particle processes in the Universe.  These extreme sources produce non-thermal spectral energy distributions (SEDs), characterized by two broad peaks.  The origin of the lower-energy peak is relatively well understood, resulting from the synchrotron radiation of relativistic leptons in the presence of a tangled magnetic field.   The higher-energy SED peak is commonly attributed to inverse-Compton up-scattering by the relativistic leptons within the jet of either the synchrotron photons themselves, namely synchrotron self-Compton (SSC) emission, or a photon field external to the jet, namely external Compton (EC) emission.   Alternative models attribute the higher-energy peak of blazar emission to hadronic pion production and the resulting cascade emission, which can provide convincing evidence that sufficient hadronic acceleration is at work within VHE blazars to make them reasonable source of UHECRs.

VHE gamma rays that propagate through the intergalactic medium are absorbed by low energy extragalactic background light (EBL) photons via pair production, $\gamma + \gamma \rightarrow e^{+} + e^{-}$\cite{nikishov}.  The absorption modifies the intrinsic VHE gamma-ray spectra of extragalactic objects and limits the distance out to which these sources should be detectable by VHE instruments.  The modification of the emitted spectrum is energy and redshift-dependent, making the distance to extragalactic VHE sources a vital parameter for the accurate interpretation of the observed VHE spectra.

The absorption of VHE gamma rays by the EBL is estimated using the model-specific (e.g. \cite{gilmore, dominguez, franceschini} gamma-ray opacity, $\tau(E,z)$, and the intrinsically emitted flux, $F_{int}$, is estimated from the observed flux, $F_{obs}$, using the relation $F_{int} = F_{obs}\times e^{\tau(E,z)}$.  Although the EBL cannot be directly measured due to strong foreground sources, alterations to intrinsically emitted VHE blazar spectra by absorption by the EBL have been used to estimate the spectral properties of the EBL \cite{aharonian2006,albert2008}, providing upper limits on the IR photon density which are consistent with the strict observational lower limits set by galaxy counts.  

Recent work has indicated that the EBL density is closer to the observational lower limits than the indirectly set upper limits \cite{HESSEBL,fermiEBL}.  Interestingly, there are VHE observations which, when corrected for absorption by even the lowest density EBL models, show indications of spectral hardening at the highest energies, but are strictly in agreement with the $\Gamma=1.5$ spectral limitation (where $dN/dE \propto E^{-\Gamma}$) described in, e.g., \cite{aharonian2006}.

The onset of spectral hardening at the highest energies in sources which probe moderate opacities is clearly predicted by theories which include secondary emission from UHECR extragalactic propagation and interaction with intervening diffuse photon fields such as the EBL and CMB.    If VHE blazars eject sufficiently accelerated protons, photo-pion production with these photon fields can initiate cascades along the line of sight, producing a hard spectral feature at the highest energies from secondary gamma rays, as described in \cite{essey2010}.  Detected gamma-ray emission can be definitively associated with these secondary emission processes using the spectral shape and variability characteristics that are observed in the gamma-ray band.  For distant sources, it is expected that the secondary component would contribute a significant portion of the observed gamma-ray signal, particularly at the highest energies, where intrinsic gamma rays are attenuated over the long path length through the EBL, as explored in \cite{essey2014}.   Any hard component displayed by extragalactic VHE sources should lack variability due to intrinsic variations in flux being washed out over a variety of path lengths - a useful observable for checking the feasibility of the secondary line-of-sight UHECR emission scenario.

Studying VHE blazars also enables the investigation of the magnitude of the intergalactic magnetic field (IGMF) through the study of the cascade emission along the source line of sight.   The large scale magnetic fields that may exist in the regions between galaxies is known as the IGMF.  Although the origin of this magnetic field is still unknown, and could have been generated at several points in the evolution of the early universe \cite{voids2}, the current IGMF has been proposed as the seed field for strong ($10^{-6}$ G) fields observed in galaxies and galaxy clusters.  Recent studies indirectly place upper ($10^{-9}$ G) and lower ($10^{-20}$ G) limits on the strength of the IGMF \cite{EGMFbound1, EGMFbound2}.  Tighter constraints, and possible insight on any large scale structure (e.g. correlation length) on the field strength are necessary to constrain models explaining the IGMFÕs origin.  The inclusion of this affect in emission models can enable an indirect probe to the IGMF through the inspection of extragalactic gamma-ray point source diffusion beyond the instrumental PSF.   There have also been studies on the possibility that for bright blazars the pair cascade energy is dissipated in heating of the intergalactic medium \cite{broderick, schlickeiser}, a process that may need to be taken into account when using VHE blazars to study the IGMF and a topic where observations of bright VHE blazars which are significantly attenuated by EBL absorption can provide insight.

Notably, there is an additional, more exotic, mechanism potentially affecting gamma-ray propagation over extragalactic distances.  A hard high-energy gamma-ray tail similar to that produced from UHECR interactions (or over-estimated absorption by the EBL) would result if VHE photons oscillate into axion-like particles (ALPs), allowing propagation through the EBL without interaction \cite{deAngelis2011}.  Evidence for an effect of just this sort has recently been claimed \cite{meyer}. 

Motivated by the potential studies of extragalactic gamma-ray photon propogation and the interacting cosmological fields, VERITAS has collected deep observations of three distant VHE blazars (PKS 1424+240 at $z>0.6035$, PG 1553+113 at $z>0.395$ and 3C 66A at $0.3341 < z < 0.41$).  We discuss the VERITAS observations, source characteristics and the flux correction for absorption by the EBL here.

\begin{figure}[h!]
\centering
\includegraphics[width=75mm]{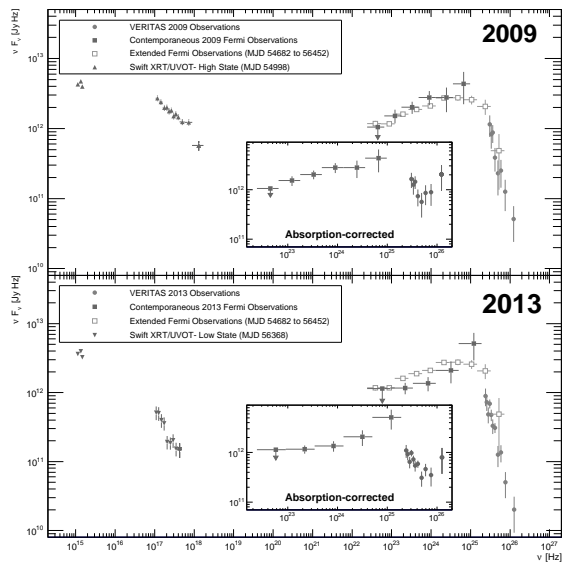}
\caption{Two broadband SEDs of PKS\,1424+240, corresponding to a relatively high (upper panel) and a low (lower panel) state. Within the inset, the VHE data are corrected for absorption using the low-opacity Gilmore et al. 2012 EBL model for $z=0.6$.  The LAT data above 100 MeV, contemporaneous with the VERITAS observations in each year are shown along with the spectral results from full LAT observations. The \textit{Swift} XRT and UVOT observations of a relatively low and high state are also shown.} \label{Figure1424}
\end{figure}

\section{VERITAS Observations of Distant VHE Blazars}
\subsection{PKS 1424+240}
PKS\,1424+240 (VER\,J1427+237) is a distant very high energy (VHE; $E\ge100$ GeV) blazar at $z\ge0.6035$ \cite{furniss1424}.  At this \textit{minimum} distance, the intrinsic VHE emission is significantly absorbed by the EBL.  VERITAS observations of PKS\,1424+240 were performed over three seasons and are reported in \cite{new1424}.  The first season (MJD 54881-55003) provides 28 hours of quality-selected livetime and is reanalyzed here, showing results consistent with those reported in \cite{acciari1424}.  The second season encompasses 14 quality-selected hours of observation between MJD 55598 and 55711, while the third season includes data spanning MJD 56334 to 56447, and provide 67 hours of quality-selected livetime with a threshold of 100 GeV, enabled by a camera upgrade in 2012.  

The contemporaneous broadband SEDs of PKS 1424+240 for the relatively high state observed in 2009 and the relatively low state of 2013 are shown in Figure 1.  When the observed VHE spectrum of PKS\,1424+240 is corrected for the minimal absorption by the EBL, it appears that the source displays a complex spectral structure (insets of Figure 1).   With the marginal ($\sim2\sigma$) hardening at the highest energies, it is challenging to model the source emission with a standard SSC emission scenario.

\begin{figure}[h!]
\centering
\includegraphics[width=75mm]{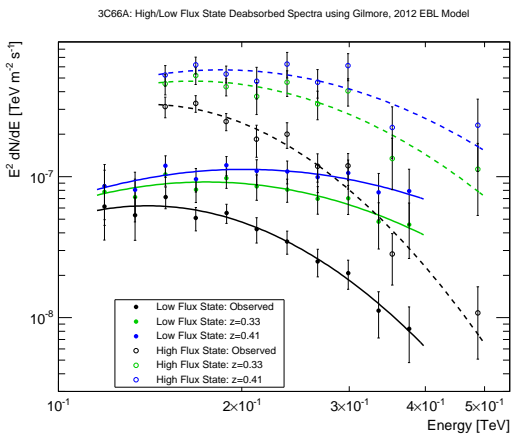}
\caption{VHE spectra of 3C 66A in both the low (solid black) and high (dotted black) flux state.  The elevated state was observed over three days in October of 2008 (6 hours of livetime; MJD 54747-54749), while the low state is results from integrated observations between 2008 and 2014 (52 hours of livetime; MJDs 54361-54746 \& 54750-56604).  The observations are corrected for the expected absorption by the \cite{gilmore} EBL model, assuming the upper (blue) and lower (green) redshift limits of $z=0.41$ and $z=0.33$, respectively. } \label{Figure3C66A}
\end{figure}

\subsection{3C 66A}
As a bright gamma-ray source which was for a long time purported to be at a redshift of $0.44$ from a erroneous spectroscopy measurements \cite{miller,kinney, bramel}, 3C\,66A is of great interest to the high-energy astrophysics community.  Motivated by the featureless optical spectra displayed by the source, far-ultraviolet spectra were collected with COS onboard HST.  Intergalactic medium absorption features within the spectra were used to place a firm lower limit on the blazar redshift of $z\ge0.3347$.  Additionally, an upper limit is set by statistically treating the non-detection of additional absorbers beyond $z=0.3347$, indicating a redshift of less than 0.41 at 99\% confidence and ruling out $z\ge$0.444 at 99.9\% confidence \cite{furniss3C66A}.  

This BL Lac object was first detected at VHE by VERITAS in 2008 \cite{3C66Apaper,3C66AMWL}.  VERITAS observations in October of 2008 showed the source to be in an elevated state, displaying VHE emission at the level of 6\% of the Crab (6 hours of livetime between MJD 54747-54749; dashed black line in Figure 2).   Continued observation of the source were completed between 2008 and 2014, resulting in 52 hours of exposure while the source displayed a lower flux state of $\sim$2\% of the Crab Nebular (MJDs 54361-54746 \& 54750-56604; solid black line Figure 2).    The high and low states displayed by 3C\,66A can both be fit with a curved log parabola of the form $E^2 dN/dE = N_0 (E/E_0)^{-\Gamma-\beta {\rm log}(E/E_0)}$ (with $E_0$= 270 GeV) where $N_0=(11.8\pm1.3) \times 10^{-8}$ and $(2.82\pm0.28) \times 10^{-8}$ TeV m$^{-2}$ s$^{-1}$ , $\Gamma=3.07\pm0.4$ and $2.30\pm0.43$ and $\beta=5.2\pm2.1$ and $4.8\pm1.8$ for the high and low states, respectively.  The log-parabolic fit is preferred over a simple power-law fit at the $\sim3\sigma$ level for both the low and high states.

With detection of 3C 66A out to 400 GeV in the low state, and 500 GeV in the high state, the gamma-ray opacity being probed by the source can be estimated for the redshift lower and upper limits according to the low density EBL model from \cite{gilmore}.  For the low state, $\tau(z=0.33, 400$ GeV)=1.4 and $\tau(z=0.41, 400$ GeV)=1.8.  For the high state,  $\tau(z=0.33, 500$ GeV)=1.8 and $\tau(z=0.41, 500$ GeV)=2.5.

\begin{figure*}[t!]
\centering
\vspace{-30ex}
\includegraphics[width=75mm]{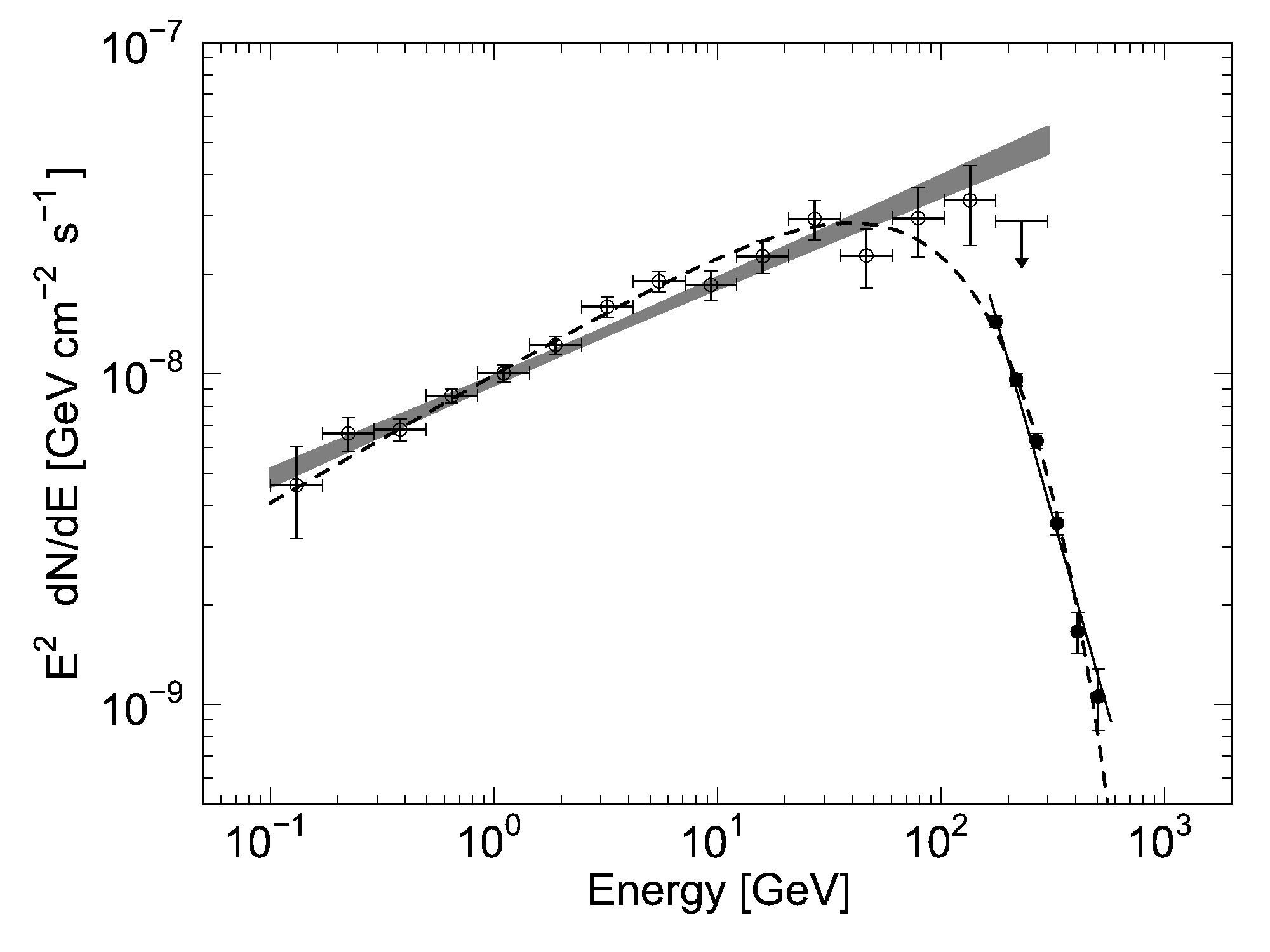}\includegraphics[width=75mm]{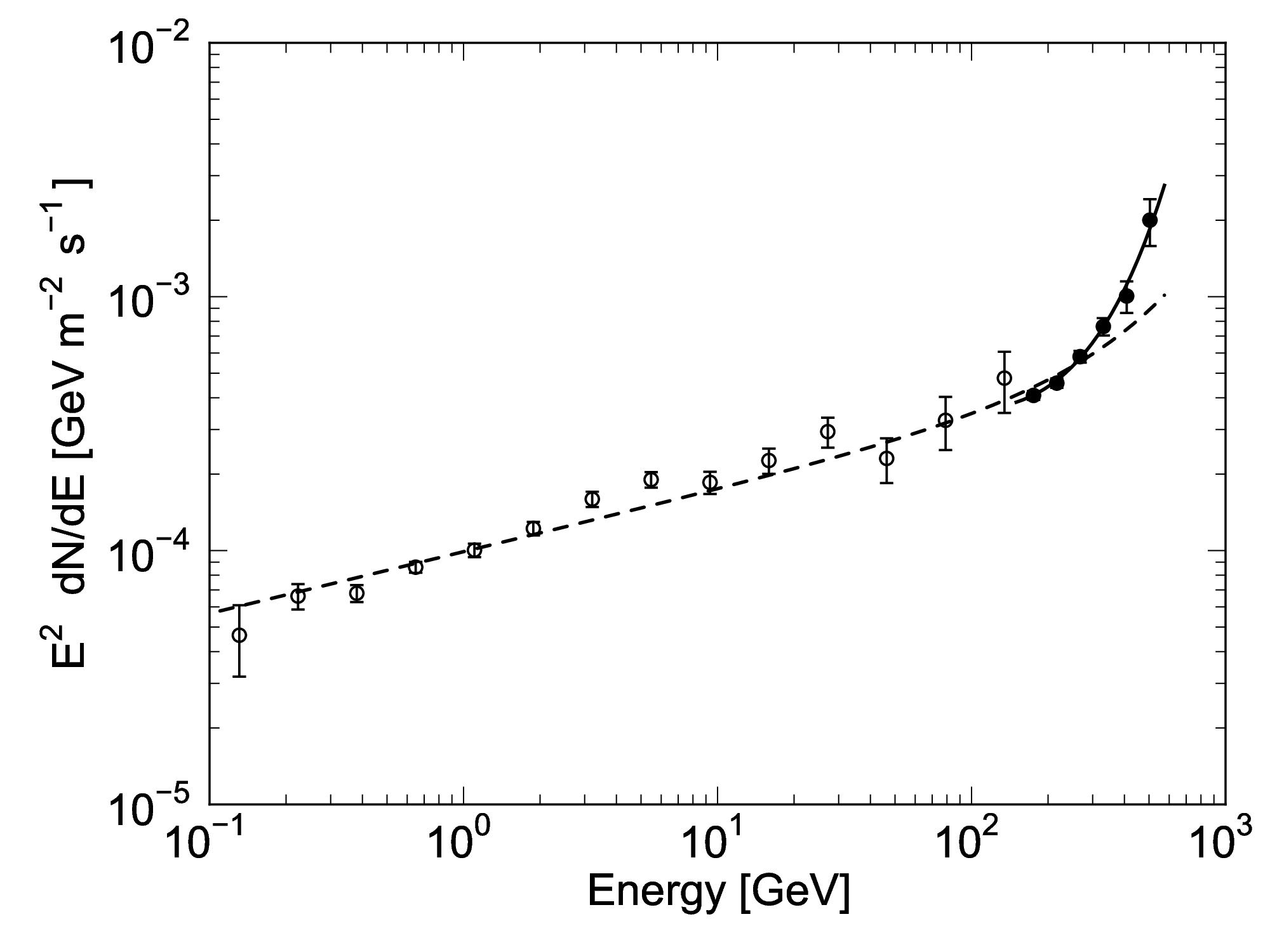}
\caption{\textbf{Left:} $Fermi$-LAT spectrum of PG 1553+113 (grey shaded area and open data points) plotted along with the VERITAS-observed spectrum (solid black data points and line). The highest-energy Fermi-LAT spectral bin is shown with a 95\% confidence level upper limit. The best fit to the combined spectrum using a power law with an exponential cutoff are shown with a dashed line.  \textbf{Right:} The absorption-corrected VHE spectrum (filled circle data points) a redshift of $z=0.53$ and the EBL model from \cite{gilmore}. The solid curve represents the best fit to the intrinsic VHE spectrum using a power law with an exponential rise and was the fit used to set the upper limit on the source redshift. The dashed curve shows the best fit to the absorption-corrected gamma-ray spectrum.  } \label{Figure1553}
\end{figure*}

\subsection{PG 1553+113}
PG\,1553+113 is readily detected in the high-energy (100 MeV to 100 GeV) and VHE gamma-ray regimes.    VERITAS is capable of detecting PG 1553+113 above 100 GeV with a significance of 5$\sigma$ after $\sim$43 minutes of exposure, given its average flux of 6.9 \% Crab.  Without a firm spectroscopic redshift due to a featureless optical spectrum, recent UV measurements using the COS on HST have yielded the strictest redshift constraint on the source to date, setting a firm lower limit of $z \ge 0.395$ \cite{danforth}.  

The observed VHE spectrum displayed in Figure 3 (left panel) results from 95 hours of observation, and is shown along with contemporaneous $Fermi$ observations.  These observations are detailed in \cite{1553paper}.  The VERITAS spectrum is measured between 160 and 560 GeV, and is well defined by a differential power law with index $4.33\pm0.09$.  The combined contemporaneous $Fermi$ LAT and VERITAS data between 100 MeV and 560 GeV are well fit with a power-law with an exponential cutoff at 101.9$\pm$3.2 GeV.  Part of the cutoff at this energy may be instrinsic, but it is expected that a significant fraction of the cutoff is due to absorption by the EBL.   Through EBL absorption-correction using the model from \cite{gilmore}, and the physically motivated requirement that the gamma-ray spectrum does not display an intrinsic exponential rise, the VHE spectrum measured by VERITAS allows a robust upper limit on the distance to the source of $z\le0.62$ (Figure 3, right panel).

\section{Conclusions}
PKS 1424+240,  3C 66A and PG 1553+113 are all relatively bright VHE emitting blazars at significant redshifts.  The bright VHE emission and significant distances have motivated deep VERITAS observations over multiple years .  The steady emission allows spectral reconstruction up to hundreds of GeV,  making the sources good targets for the study of VHE photon interaction with EBL photons along the line of sight.  Improvement of our understanding of VHE blazar emission, VHE photon propagation across extragalactic distances and the cosmoligical fields which VHE photons interact with will continue to motivate VERITAS observations of these unique sources.

\bigskip 
\begin{acknowledgments}
This research is supported by grants from the U.S. Department of Energy Office of Science, the U.S. National Science Foundation and the Smithsonian Institution, by NSERC in Canada, by Science Foundation Ireland (SFI 10/RFP/AST2748) and by STFC in the U.K. We acknowledge the excellent work of the technical support staff at the Fred Lawrence Whipple Observatory and at the collaborating institutions in the construction and operation of the instrument.
\end{acknowledgments}

\end{document}